\documentclass[conference]{IEEEtran}
\IEEEoverridecommandlockouts
\usepackage{cite}
\usepackage{algorithmic}
\usepackage{graphicx}
\usepackage{textcomp}
\usepackage{xcolor}
\usepackage{soul}
\usepackage{amssymb,amsmath,amsthm,enumitem}
\usepackage{graphicx}
\usepackage{csquotes}
\usepackage{float}
\usepackage{gensymb}
\usepackage[labelfont=bf,figurename=Fig.]{caption}
\usepackage{tabu}
\usepackage{mathtools}
\usepackage{csvsimple}
\usepackage{array,booktabs}
\usepackage{textcomp}
\usepackage{epsfig}
\usepackage{setspace}

\usepackage{makecell}
\usepackage{multirow}
\usepackage{textgreek}
\usepackage{wasysym}
\usepackage{sidecap}

\usepackage{booktabs}

\usepackage{upgreek}
\usepackage{nameref}
\usepackage{soul}

\def\BibTeX{{\rm B\kern-.05em{\sc i\kern-.025em b}\kern-.08em
    T\kern-.1667em\lower.7ex\hbox{E}\kern-.125emX}}
\begin{document}

\title{Synthetic ultrasound images to benchmark echocardiography-based biomechanics\\
\thanks{This work was supported by the National Institutes of Health R00HL138288 to R.A., and the American Heart Association predoctoral fellowship 24PRE1240097 to T.M.}
}

% \author{\IEEEauthorblockN{Tanmay Mukherjee}
% \IEEEauthorblockA{\textit{Department of Biomedical Engineering} \\
% \textit{Texas A\&M University}\\
% College Station, TX, USA \\
% Email: tanmaymu@tamu.edu}
% \and
% \IEEEauthorblockN{Sunder Neelakantan}
% \IEEEauthorblockA{\textit{Department of Biomedical Engineering} \\
% \textit{Texas A\&M University}\\
% College Station, TX, USA \\
% Email: sundern@tamu.edu}
% \and
% \IEEEauthorblockN{Kyle Myers}
% \IEEEauthorblockA{\textit{Hagler Institute for Advanced Study} \\
% \textit{Texas A\&M University}\\
% College Station, TX, USA \\
% Email: drkylejmyers@gmail.com}
% \and
% \IEEEauthorblockN{Carl Tong}
% \IEEEauthorblockA{\textit{College of Medicine} \\
% \textit{Texas A\&M University}\\
% College Station, TX, USA \\
% Email: ctong@tamu.edu}
% \and
% \IEEEauthorblockN{Reza Avazmohammadi}
% \IEEEauthorblockA{\textit{Department of Biomedical Engineering} \\
% % \textit{J. Mike Walker ’66 Department of Mechanical Engineering} \\
% \textit{Texas A\&M University}\\
% College Station, TX, USA \\
% Email: rezaavaz@tamu.edu}
% }

\author{
\IEEEauthorblockN{Tanmay Mukherjee\IEEEauthorrefmark{1},
Sunder Neelakantan\IEEEauthorrefmark{1},
Kyle Myers\IEEEauthorrefmark{2},
Carl Tong\IEEEauthorrefmark{3} and
Reza Avazmohammadi\IEEEauthorrefmark{1}\IEEEauthorrefmark{4}}
\IEEEauthorblockA{\IEEEauthorrefmark{1}Department of Biomedical Engineering, 
Texas A\&M University, College Station, TX, USA\\}
\IEEEauthorblockA{\IEEEauthorrefmark{2}Hagler Institute for Advanced Study, Texas A\&M University, College Station, TX, USA\\}
\IEEEauthorblockA{\IEEEauthorrefmark{3}College of Medicine, Texas A\&M University, College Station, TX, USA\\}
\IEEEauthorblockA{\IEEEauthorrefmark{4}J. Mike Walker ’66 Department of Mechanical Engineering, 
Texas A\&M University, College Station, TX, USA\\ Email: rezaavaz@tamu.edu}}

\maketitle

\begin{abstract}
Brightness mode (B-mode) ultrasound is a common imaging modality in the clinical assessment of several cardiovascular diseases. The utility of ultrasound-based functional indices such as the ejection fraction (EF) and stroke volume (SV) is widely described in diagnosing advanced-stage cardiovascular diseases. Additionally, structural indices obtained through the analysis of cardiac motion have been found to be important in the early-stage assessment of structural heart diseases, such as hypertrophic cardiomyopathy and myocardial infarction. Estimating heterogeneous variations in cardiac motion through B-mode ultrasound imaging is a crucial component of patient care. Despite the benefits of such imaging techniques, motion estimation algorithms are susceptible to variability between vendors due to the lack of benchmark motion quantities. In contrast, finite element (FE) simulations of cardiac biomechanics leverage well-established constitutive models of the myocardium to ensure reproducibility. In this study, we developed a methodology to create synthetic B-mode ultrasound images from FE simulations. The proposed methodology provides a detailed representation of displacements and strains under complex mouse-specific loading protocols of the LV.  A comparison between the synthetic images and FE simulations revealed qualitative similarity in displacement patterns, thereby yielding benchmark quantities to improve the reproducibility of motion estimation algorithms. Thus, the study provides a methodology to create an extensive repository of images describing complex motion patterns to facilitate the enhanced reproducibility of cardiac motion analysis.

\end{abstract}


%%%%%%%%%%%%%%%%%%%%%%%%
% INTRODUCTION
%%%%%%%%%%%%%%%%%%%%%%%%
\section{Introduction}
% carry over relevant suggestions from the abstract to the same sentences here
Clinical assessment of several cardiovascular diseases is performed through cine imaging via brightness mode (B-mode) ultrasound \cite{Marwick-2006, Marwick-2018}. The utility of functional indices such as the ejection fraction (EF) and stroke volume (SV) is widely described in diagnosing advanced-stage cardiovascular diseases \cite{Mewton-2013, Wehner-2020}. Additionally, structural indices obtained through the analysis of cardiac motion, such as the global circumferential and longitudinal strains (GCS and GLS, respectively), are sensitive to systolic dysfunction and have found importance in diagnosing early-stage pathology \cite{Flores-Ramirez-2017}. B-mode imaging, in conjunction with speckle tracking, remains the most widely implemented methodology in quantifying such structural indices. Despite the promise of structural indices in advancing the kinematic understanding of cardiac motion, the spatially heterogeneous strain distribution, influenced by anisotropy in the microstructure of the heart, may challenge reproducibility. This complexity is manifested in the variability of regional and global measures exhibited among medical imaging vendors and operators \cite{Pedrizzetti-2016}, which is further complicated by the absence of definitive "ground-truth" strain values. 

% heart, current ultrasound methods fall short in describing the regional variations in strain patterns. Indeed

\begin{figure}[ht!]
\centering
\includegraphics[width=0.95\columnwidth]{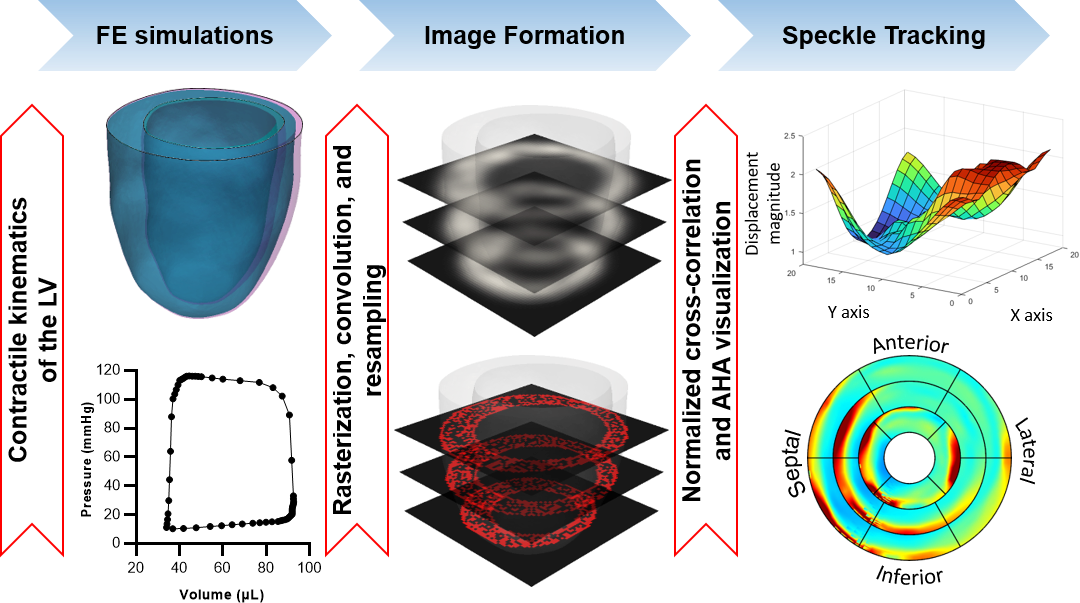}
\caption{Schematic outlining the stages in creating synthetic B-mode ultrasound images from finite element simulations of cardiac kinematics.}
\label{fig:schematic}
\end{figure}

In contrast, \textit{in-silico} finite element (FE) simulations that leverage subject-specific geometries, myocardial architecture, and pressure distributions have enhanced the mechanical characterization of the cardiac tissue \cite{AvazABME-2018, AvazANNUREV-2019}. Such biomechanical models of the heart have shown promise in describing pathophysiological conditions by establishing a kinematic benchmark. The incorporation of biomechanical models with imaging has been constrained to tasks such as image segmentation and enhancement \cite{Suinesiaputra-2016}, with motion analysis yet to be fully integrated. In this study, we developed a methodology to create synthetic B-mode ultrasound images from FE simulations of the heart and perform speckle tracking (Fig. \ref{fig:schematic}). Additionally, we analyzed the sensitivity of image formation to the detection of the spatial heterogeneity in cardiac motion. In essence, we facilitate the creation of a database of ultrasound images with ground-truth motion estimation to enhance the reproducibility of regional strain assessments in cardiac health and disease.

%%%%%%%%%%%%%%%%%%%%%%%%
% METHODS
%%%%%%%%%%%%%%%%%%%%%%%%
\newcommand\norm[1]{\lVert#1\rVert}

\setlength{\belowcaptionskip}{-20pt}

\section{Methodology}
\subsection{FE simulations of a mouse-specific LV}
The kinematic behavior of the heart over the course of the cardiac cycle was modeled via FE simulation of a mouse-specific LV model. A  mouse-specific geometry was subjected to a pressure-volume loop obtained through catheterization. The geometry was simulated for a full cardiac cycle, and the resulting data has been discussed extensively in previous studies \cite{AvazABME-2018, Mendiola-2022}. Briefly, the heart geometry was reconstructed using cardiac magnetic resonance images and subsequently meshed in Materialize 3-Matic. The space between the two boundaries, i.e., endo- and epicardium, was filled through an adaptive mesh strategy, with the minimum element size maintained at 0.5 mm. The distance between nodes at the boundaries was specified at 1 mm, measured as the length of the hypotenuse of a tetrahedral element. Next, myofibers were mapped onto the meshed geometry using a Laplace-Dirichlet rule-based algorithm \cite{AvazABME-2018}. The LV was divided into six layers between the endo- and epicardium, and the helicity ($\theta$) of the fibers was allowed to range from $\mathrm{\theta_{endo} = 84\degree}$ to $\mathrm{\theta_{epi} = -42\degree}$ for the mouse-specific LV. The myocardium was modeled as a transversely isotropic material with the myofiber direction ${\bf N}$ and the hyperelastic energy function $W$ characterizing the passive part. The Cauchy stress $T$ is then decomposed as: 
\begin{equation} \label{DecompStr}
\bf{T} = \underbrace{\frac{1}{J}\, \bar{\bf F}\,\frac{\partial W^{dev}}{\partial \bar{\bf E}}\, \bar{\bf F}^{T} +\frac{\partial W^{vol}}{\partial J}} _ {\textrm{Passive}} + \underbrace{ \frac{1}{J}\, {\bf F}\,{\bf S}^{act}\, {\bf F}^{T}}_{\textrm{Active}},
\end{equation}
where \textbf{F} is the deformation gradient, \textit{J} denotes the deformation volumetric changes, and deviatoric part of $\bf F$ is represented by $\bar{\bf F} = J^{-1/3}\bf F$. Deviatoric and volumetric components were represented by $W^{dev}$ and $W^{vol}$, respectively, and \textbf{E} is the Green-Lagrange strain tensor. 
The second Piola–Kirchhoff active stress tensor is given by
\begin{equation} \label{Sact}
{\bf S}^{act}=\frac{T_a ({E}_f)}{2{E}_f+1}\,{\bf N} \otimes {\bf N},
\end{equation}
where $T_a ({E_f})$ is a stress-like positive function of the strain in the fiber direction ${\bf N}$ given by ${E}_f= {\bf N}\cdot {\bf E} {\bf N}$. We chose {\color{black}the following form for $T_a ({E}_f)$} 
\begin{equation} \label{Sa}
T_a ({E}_f) = T_{\rm{Ca}^{2+}}\,\left[1+\beta\,\left(\sqrt{2 {E}_f+1} - 1 \right) \right].
\end{equation}
In this model, the active force $T_{\rm{Ca}^{2+}}$, generated for the resting myofibers, increases by a positive factor $\beta$ when the myofibers are extended to the strain ${E}_f$, obeying the Frank-Starling relationship.

\subsection{Strain calculations}
Following in-silico simulations, the large deformation formulation was used to derive the Green-Lagrange strain tensor ($\mathbf{E}$). The elemental FE strains were projected onto the centroids, such that strains were obtained as pointwise data. Strains were derived using the total deformation gradient ($\mathbf{F}$) and the identity matrix ($\mathbf{I}$) as:
\begin{equation}\label{eq:cart_strain}
\bf E \rm = \frac{1}{2} \left( \bf F^{\rm T} \bf F \rm - \bf I \right), 
\end{equation}
where $\mathbf{F}$ is the deformation gradient calculated as: 
\begin{equation}\label{eq:deform_gradient} 
\bf F = \bf I \rm + \frac{\partial \bf u}{\rm \partial \bf X} \hspace{0.1in} = 
\left[ 
\begin{matrix}
\frac{\partial u_{x}}{\partial X} & \frac{\partial u_{x}}{\partial Y} & \frac{\partial u_{x}}{\partial Z} \\
\frac{\partial u_{y}}{\partial X} & \frac{\partial u_{y}}{\partial Y} & \frac{\partial u_{y}}{\partial Y} \\
\frac{\partial u_{z}}{\partial X} & \frac{\partial u_{z}}{\partial Y} & \frac{\partial u_{z}}{\partial Z}\\
\end{matrix}
\right],
\end{equation}
where $u_{x}$, $u_{y}$, and $u_{z}$ are the displacements in the x,y, and z directions between two consecutive time frames. The resulting Cartesian strains were converted to the widely used radial-circumferential-longitudinal (RCL) axes using an orthonormal transformation matrix ($\mathbf{Q}$) as:
\begin{equation}\label{eq:FE_cart_strain}
\bf E_{\rm [R,C,L]} \rm =  \bf QEQ^{\rm T} \hspace{0.1in} = 
\left[ 
\begin{matrix} 
E_{RR} & E_{RC} & E_{RL} \\
E_{CR} & E_{CC} & E_{CL} \\
E_{LR} & E_{LC} & E_{LL}
\end{matrix} \right],
\end{equation}
where $E_{RR}$, $E_{CC}$, and $E_{LL}$ are the radial, circumferential, and longitudinal strains, respectively.

\subsection{Creation of synthetic B-mode images}
\begin{figure}[ht!]
\centering
\includegraphics[width=0.85\columnwidth]{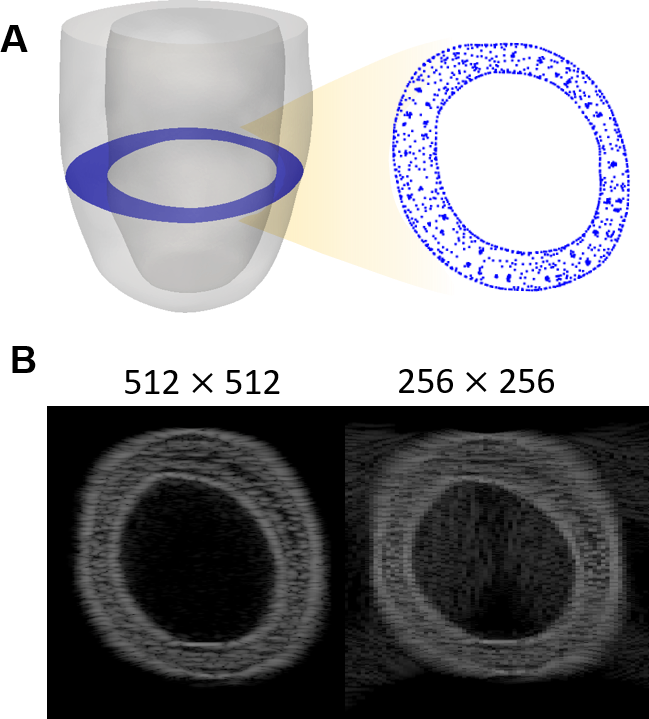}
\caption{(A) Mouse-specific left ventricle geometry, with the midsection highlighted in blue. (B) Synthetic ultrasound images were created through rasterization of the finite element simulations of the geometry during a cardiac cycle followed by a method simulating standard B-mode phased linear array imaging. The images were generated through scanning, scattering, and apodization by adopting an acoustic wave impulse response. The original image of (left) 512 $\times$ 512 pixels was also downsampled to (right) 256 $\times$ 256 pixels.}
\label{fig:synth_img}
\end{figure}

The connectivity data for the geometries were used to create an unstructured grid of points in their respective undeformed states. The FE simulation-derived displacement vectors were used to update the nodal positions through each time increment, thus providing synthetic phantom data at various time points in the cardiac cycle. Each unstructured grid or phantom ($\bf X \rm \in \mathbb{R}^{\rm 2} \rm (x,y)$) was then sectioned into three short-axis (SA) planes (base, mid, and apical) and one long-axis (LA) plane. Synthetic B-mode images were created through the Field II Ultrasound Simulation Program \cite{Jensen-2004}. The transducer was modeled as a phased linear array with 64 active transducer elements, and the impulse response of the array was modeled as a sinusoidal wave  calculated as the solution to the acoustic wave equation as:
\begin{equation}\label{eq:img_form}
\frac{\partial^{2}\cal H}{ \rm \partial \it t^{2}} =  \it c^{\rm 2} \frac{\partial^{2}\cal H}{ \rm \partial \it \bf X^{\rm \it 2}},
\end{equation}
\begin{equation}\label{eq:img_kernel}
\cal H \rm (x,y) = A \, sin(2 \pi \hspace{.1em} f \hspace{.1em}x + \phi) + B \, cos (2 \pi \hspace{.1em} f\hspace{.1em} y + \phi),
\end{equation}
where c is the wave speed, f is the wave number, $\phi$ is the phase, and A and B are the wave amplitudes in the x and y directions, respectively. The response amplitude was determined to be a normal distribution, with apodization performed using the built-in functions of Field II, and the resulting voltage trace was converted to 8-bit images of a fixed resolution of 512 $\times$ 512 pixels. Furthermore, the images were downsampled to predefined sizes of 128 $\times$ 128 and 256 $\times$ 256 to perform different standards of motion calculations. Here, the $512 \times 512$ image corresponds to a resolution of 1 px/mm, whereas the $256 \times 256$ describes 0.5 px/mm. Given the rapidness of the wall motion and the fineness of the mesh describing the geometry, the synthetic images presented different sources of artifacts. The images created from the FE simulations were presented at the mid-slice of the mouse-specific geometry for three sets of spatial resolution (Fig. \ref{fig:synth_img}A).

\setlength{\belowcaptionskip}{-10pt}

\subsection{Speckle tracking}
A speckle-tracking framework was implemented to evaluate cardiac strains at end-systole (ES) with respect to end-diastole (ED) \cite{MukherjeeSPIE-2023}. A normalized cross-correlation algorithm was used to track pixel movement between two time-dependent interrogation windows, namely, $\mathbf{w}_{{1}}$ and $\mathbf{w}_{{2}}$, in the ${k^{th}}$ Fourier space as: 
\newcommand*\conj[1]{\overline{#1}}
\begin{align}\label{equ: NCC}
\mathbf{u} =  
% \underset{\mathbf{u}}
{\text{argmax}} \left\{ \mathcal{F}^{\, -1} \left( \frac{\mathbf{W}_{1}^{k} \, \odot \, \conj{\mathbf{W}_{2}^{k}}}{\vert \mathbf{W}_{1}^{k} \, \odot \, \conj{\mathbf{W}_{2}^{k}} \vert} \right) \right\} ,
\end{align}
where $\mathbf{u}$ is the Cartesian displacement vector, $\mathcal{F}^{\, -1}$ denotes the inverse Fourier transform, $\odot$ denotes element-wise multiplication, the overline denotes the complex conjugate, and $\mathbf{W}_{\mathrm{1}}$ and $\mathrm{\conj{\mathbf{W}}_{\mathrm{2}}}$ are the Fourier transforms of $\mathbf{w}_{\mathrm{1}}$ and $\mathrm{\conj{\mathbf{w}}_{\mathrm{2}}}$, respectively. Thus, $\mathbf{u}$ was determined as the location of the peak normalized-cross correlation between $\mathbf{w}_{{1}}$ and $\mathbf{w}_{{2}}$, respectively. Interrogation windows of 64 x 64 pixels were set up initially, and the resulting displacements were linearly interpolated onto a grid space of 10 x 10 pixels. 
% A spline smoothing function was used to reduce outliers and noise. 
% The Cartesian displacements were then used to derive the Green-Lagrange strain tensor ($\mathbf{E}$). 
After calculating Cartesian displacement vectors, they were converted to the polar directions. Subsequently, the accuracy of the calculations was estimated by comparing the root mean squared (RMS) error between FE- and image-derived displacements. 

%%%%%%%%%%%%%%%%%%%%%%%%
% RESULTS
%%%%%%%%%%%%%%%%%%%%%%%%

\section{Results}
\begin{figure}[ht!]
\centering
\includegraphics[width=0.9\columnwidth]{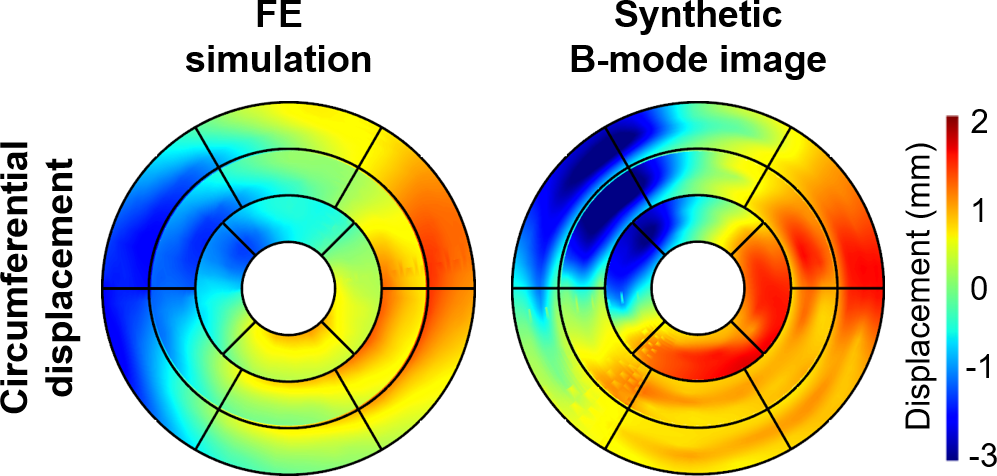}
\caption{Representative distribution of the circumferential displacement vectors obtained via finite element simulations and speckle tracking in the $512 \times 512$ synthetic images shown as an AHA bullseye visualization map. The outermost circle corresponds to the base, and the innermost corresponds to the apical slice of the left ventricle.}
\label{fig:aha_disp}
\end{figure}

The FE simulations indicated significant circumferential contraction ($E_{CC} = -0.1825 \pm 0.0315$) and wall thickening ($E_{RR} = 0.2411 \pm 0.0407$) and longitudinal shortening in the mouse-specific LV geometry. Speckle tracking provided insight into the similarities in motion obtained from the FE simulations, and results are presented for the displacement of the heart at ES with respect to ED. The corresponding polar displacement vectors were presented as standard American Heart Association (AHA) bullseye visualization maps (Fig. \ref{fig:aha_disp}). Despite subtle differences in the absolute value of the displacements, the contractile behavior of the LV was qualitatively captured using speckle tracking. However, a more pronounced decline in the transmural spread of the displacement (measured by the standard deviation of the displacement distribution) from the endo- to epicardium was observed in the FE data. Whereas a 50\% change in the transmural spread of circumferential displacement was observed in the FE simulations, the maximum spread variation across the thickness was contained to 10\% in the image-derived displacements.  Although the mean error between the FE- and image-derived displacements in the 512 $\times$ 512 images was restricted to 20\% in the mid slice of the LV (Fig. \ref{fig:disp_error}), lowering the resolution substantially affected speckle tracking. Indeed, the rapid motion of the LV, which manifests as sub-pixel movements in low-resolution images, may not be captured through normalized cross-correlation. Despite these variations in describing the spatial heterogeneity of the LV motion, the synthetic images provided similar motion patterns of wall thickening, thus highlighting the potential of the synthetic images in capturing the correlation between the LV geometry, architecture, and motion.

\setlength{\belowcaptionskip}{-10pt}

%%%%%%%%%%%%%%%%%%%%%%%%
% DISCUSSION
%%%%%%%%%%%%%%%%%%%%%%%%

\section{Discussion}
The kinematic assessment of cardiac motion has gained significant importance in the diagnostic protocol of cardiac diseases \cite{Marwick-2018}. However, a common limitation in most existing standards is variability in regional strain calculations, which are challenged by the heterogeneous and anisotropic nature of cardiac motion. The objective of this study was to provide an early benchmark to improve motion estimation through B-mode ultrasound imaging by comparing image-derived displacement with a well-known material model of the heart. Additionally, we presented a methodology to create a repository of synthetic ultrasound images from FE simulations and conducted experiments to investigate the accuracy of speckle tracking in estimating the transmural (endo- to epicardium) variations in cardiac motion. 

\setlength{\belowcaptionskip}{-10pt}
\definecolor{OliveGreen}{RGB}{85, 107, 47}
\begin{figure}[ht!]
\centering
\includegraphics[width=0.8\columnwidth]{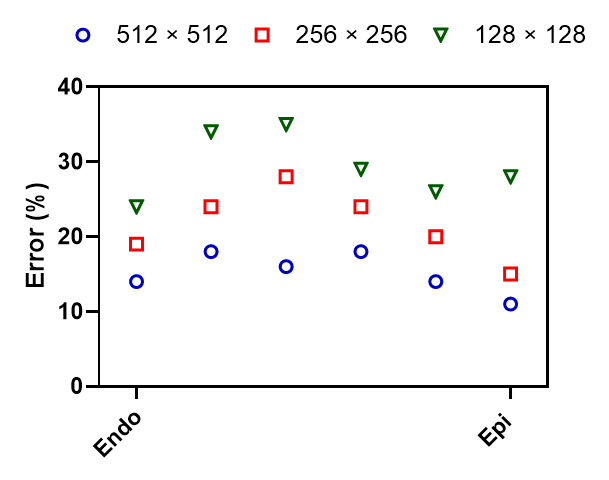}
\caption{Transmural distribution of the root mean squared error in circumferential displacements between the finite element and speckle tracking simulations. Results are presented as the RMS error at the mid and apical slice for six transmural layers between the endo- and epicardium. Images with the following spatial resolution \color{blue}$\circ$ \color{black} $512 \times 512$, \color{red}{$\Box$} \color{black}  $256 \times 256$, and \color{OliveGreen}{$\nabla$} \color{black} $128 \times 128$ pixels were used for the analyses.} 
\label{fig:disp_error}
\end{figure}

Image quality was observed to be very sensitive to the spatial heterogeneity in cardiac motion. Whereas the high-resolution images qualitatively captured the overall motion of the LV, including transmural variations in the absolute displacement magnitude, significant errors ($\sim$25\%) were noted in the lower-resolution images (Fig. \ref{fig:disp_error}). We attribute these errors to challenges in estimating sub-pixel motion due to the rapid non-affine deformation of the heart. These errors primarily highlight the utility of the proposed synthetic image repository in establishing ground truth or benchmark motion quantities to further the fidelity of image-based motion analysis. For instance, since the FE simulations facilitate the easy manipulation of material parameters, geometry, and architecture, any number of images can be synthesized to create a vast dataset in training for tasks such as classification, segmentation, and motion estimation \cite{MukherjeeSPIE-2024}.  Ultimately, the dataset serves as a foundation for training machine-learning models, enhancing the adaptability of these models across various image-processing tasks.
In particular, we anticipate the formulation of inverse models to estimate soft tissue parameters using image-based motion analysis and biomechanical modeling \cite{Babaei-2022, UsmanFIMH-2023}, thereby improving the in-vivo characterization of cardiac biomechanics.

\section*{Acknowledgment}

This work was supported by the National Institutes of Health R00HL138288 to R.A., and the American Heart Association predoctoral fellowship 24PRE1240097 to T.M.

\end{document}